\documentclass[english,PRE, twocolumn]{revtex4-1}
\usepackage[T1]{fontenc}
\usepackage[latin9]{inputenc}
\usepackage{amsmath}
\usepackage{amssymb}
\usepackage{graphicx}
\usepackage{babel}
\begin{document}
\title{Large deviation of long time average for a stochastic process : an
alternative method.}
\author{Bahram Houchmandzadeh}
\affiliation{CNRS, LIPHY, F-38000 Grenoble, France~\\
Univ. Grenoble Alpes, LIPHY, F-38000 Grenoble, France}
\begin{abstract}
We present here a simple method for computing the large deviation
of long time average for stochastic jump processes. We show that the
computation of the rate function can be reduced to that of a partial
differential equation governing the evolution of the probability generating
function. The long time limit of this equation, which in many cases
can be easily obtained, leads naturally to the rate function. 
\end{abstract}
\maketitle

\section{Introduction.}

Markov stochastic processes are at the heart of many area of science,
ranging from statistical and quantum physics to biology and economics\citep{vankampen2007stochastic,gardiner2004handbook}.
Broadly speaking, for time continuous models, a Markovian \emph{system}
in a state $\mathbf{x}$ at time $t$ will transit to a state $\mathbf{y}$
at time $t+dt$ with a probability that depends only on $\mathbf{x},\mathbf{y}$
and $t$ ; in other words the system has no memory. By following many
(ideally an infinite number) trajectories $\mathbf{x}(\tau)$, one
can construct the probability $P(\mathbf{u},t)$, \emph{i.e.} the
relative number of trajectories that pass through state $\mathbf{u}$
at time $\tau$. $P(\mathbf{u},t)$ is the fundamental quantity that
gives the most complete description of the studied system. 

During the last thirty years, large deviation theory of stochastic
processes has attracted a large number of investigation as a natural
reformulation of statistical physics (for a review, see \citep{touchette2009thelarge}).
More specifically, there has been a great interest in using the tools
of large deviation theory for studying the fluctuation of time-additive
quantities\citep{touchette2018introduction}. In this approach, an
information $\mathbf{r}(T)$ is extracted from \emph{each} trajectory
of a Markov process, and a probability density $p(\mathbf{r},T)$
is constructed for this information. Usually, one is interested only
in the long time limit ($T\rightarrow\infty)$. A most studied case
is the time average of the original variable $\mathbf{x}$: 
\begin{equation}
\bar{\mathbf{x}}(T)=\frac{1}{T}\int_{0}^{T}x(\tau)d\tau\label{eq:timeaverage-def}
\end{equation}
For jump processes that are the object of this note, the sum in (\ref{eq:timeaverage-def})
involves the usual Riemann integral. 

The purpose of the present note is to present a simple mathematical
framework for computing the probability density of such stochastic
time averaged quantities. The basic idea is to reduce the problem
at hand to the resolution of a partial differential equation and extract
it long time behavior. The mathematical tools are fairly standard
and in many cases lead simply to the desired results, as it will be
discussed below. 

This note is organized as follow : in the next section, we recall,
for self consistency, the basic concepts of discreet Markovian stochastic
processes, their large deviation theory and the main idea behind the
method we call dPGF$k$ that transforms the problem at hand into a
partial differential equation (PDE). Section \ref{sec:Applications-and-extensions}
is devoted to the application of the method to various well studied
stochastic processes and some of its extensions. The final section
is devoted to discussions of the methods limitations and conclusion.
The appendices handle the details of technical computations. 

\section{the dPGF$k$ method.}

We consider here first, for the sake of simplicity, one step Markovian
jump processes of the stochastic variable $n$ ($=0,1,...$). These
processes are described by their transition probabilities $W(n\rightarrow n\pm1)=W^{\pm}(n)$.
The probability $P(n,t)$ of being in state $n$ at time $t$ is governed
by the Master equation\citep{gardiner2004handbook,vankampen2007stochastic}
\begin{eqnarray}
\frac{\partial P(n,t)}{\partial t} & = & W^{+}(n-1)P(n-1,t)-W^{+}(n)P(n,t)\nonumber \\
 & + & W^{-}(n+1)P(n+1,t)-W^{-}(n)P(n,t)\label{eq:Master}
\end{eqnarray}
The above equation can be presented in the matricial form
\begin{equation}
\partial_{t}\left|P(t)\right\rangle ={\cal L}\left|P(t)\right\rangle \label{eq:master:matricial}
\end{equation}
where the (right, column) vector $\left|P(t)\right\rangle =(P(0,t),P(1,t),...)^{T}$
and the matrix ${\cal L}$ collects the rates of equation (\ref{eq:Master}):
\[
{\cal L}_{n\pm1}^{n}=W^{\pm}(n)\,\,\,\,;\,\,\,{\cal L}_{n}^{n}=-\left(W^{+}(n)+W^{-}(n)\right)
\]
where the upper (lower) index designates the row (column) of the matrix.
We suppose that the stochastic process has a stationary equilibrium
state independent of the initial conditions. 

The quantity of interest in this paper is the distribution of $\bar{n}(T),$
the long time average ($T\rightarrow\infty$) over a \emph{trajectory}
: 
\[
\bar{n}(T)=\frac{1}{T}\int_{0}^{T}n(t)dt
\]

The Large Deviation theory states \citep{touchette2009thelarge} that
the probability density of $\bar{n}$ is given by
\[
{\cal P}\left(\bar{n}(T)=x\right)=e^{-TI(x)}
\]
where the rate function $I(x)$ has its minimum and zero at $\mu=\left\langle n_{\text{eq}}\right\rangle $
of the original process.

To compute the rate function, which is the objective of the present
note, the Donsker-Varadhan (DV)\citep{donsker1975asymptotic,touchette2018introduction}
method consists of (i) building the \emph{tilted }matrix 
\begin{equation}
{\cal L}(k)={\cal L}+k{\cal D}\label{eq:tilted}
\end{equation}
where ${\cal D}$ is a diagonal matrix such that its elements are$\left({\cal D}\right)_{m}^{n}=n\delta_{m}^{n}$
; (ii) compute the largest eigenvalue of the above matrix: $\lambda(k)=\xi_{\text{max}}\left({\cal L}(k)\right)$
and (iii) compute the rate function $I(x)$ as the Legendre-Fenchel
transform of $\lambda(k)$ : 
\begin{equation}
I(x)=\underset{k}{\text{max}}\left(kx-\lambda(k)\right)\label{eq:legendre-def}
\end{equation}
Note that the DV method is written for the Hermitian conjugate of
${\cal L}(k)$, but these two matrices have the same eigenvalues and
therefore, without loss of generality, we use ${\cal L}(k)$ in this
article.

We can use the usual tools of the trade of stochastic processes for
the DV method. For any jump processes for example, we now that the
matrix ${\cal L}$ has a (left,row) eigenvector $\left\langle 1\right|=(1,1,....)$
with eigenvalues $\lambda=0$, as applying this vector to both sides
of equation (\ref{eq:master:matricial}) leads to
\begin{equation}
\partial_{t}\left\langle 1\right\rangle =\partial_{t}\left\langle 1|P\right\rangle =\left\langle 1|{\cal L}|P\right\rangle =0\label{eq:d1}
\end{equation}
which is another way of stating the obvious fact that $\partial_{t}\sum_{n}P(n,t)=0$.
The right eigenvector associated with $\left\langle 1\right|$ is
the equilibrium probabilities $\left|P_{\text{eq}}\right\rangle $.
Various average quantities can be computed by applying an adequate
left vector to equation (\ref{eq:master:matricial}). For example,
the evolution of the mean is given by 
\[
\partial_{t}\left\langle n|P\right\rangle =\partial_{t}\left\langle n\right\rangle =\left\langle n|{\cal L}|P\right\rangle 
\]
where $\left\langle n\right|=(0,1,2,...)$. Simple manipulations of
the sums involved show that 
\[
\partial_{t}\left\langle n\right\rangle =\left\langle W^{+}(n)-W^{-}(n)\right\rangle 
\]
The most complete information is contained in the probability generating
function (PGF)
\begin{equation}
\phi(z,t)=\left\langle z^{n}\right\rangle =\sum_{n}z^{n}P(n,t)\label{eq:PGF}
\end{equation}
 and it is straightforward to show that\citep{houchmandzadeh2010alternative}
\begin{equation}
\partial_{t}\phi(z,t)=(z-1)\left\langle z^{n}W^{+}(n)\right\rangle +(1/z-1)\left\langle z^{n}W^{-}(n)\right\rangle \label{eq:dPGF}
\end{equation}
When the transition rates are polynomials in $n$, the $\left\langle ...\right\rangle $
in the above expressions are reduced to derivatives of $\phi$. For
example (see appendix \ref{sec:The-dPGFk-equation.} for more details),
deriving expression (\ref{eq:PGF}) in respect to $z$, we have 
\[
\left\langle nz^{n}\right\rangle =z\frac{\partial\phi}{\partial z}\,\,\,;\,\,\,\left\langle n(n-1)z^{n}\right\rangle =z^{2}\frac{\partial^{2}\phi}{\partial z^{2}}
\]
Therefore, the evolution of the PGF is governed by a partial differential
equation on $\phi()$.

Now, we can embed the DV method into a similar problem by investigating
the Master-like equation 
\begin{equation}
\partial_{t}\left|P\right\rangle ={\cal L}(k)\left|P\right\rangle \label{eq:Masterlike}
\end{equation}
The vector $\left|P\right\rangle $ here is obviously not a probability
anymore when $k\ne0$, but for the purpose of computing the largest
eigenvalue and the rate function, this is of no consequence. For example,
for small $k$, we can solve equation (\ref{eq:Masterlike}) perturbatively
by setting 
\begin{equation}
\left|P(t)\right\rangle =\left(\left|P_{\text{0}}\right\rangle +k\left|P_{1}\right\rangle +k^{2}\left|P_{2}\right\rangle \right)e^{\left(k\eta_{1}+k^{2}\eta_{2}+...\right)t}\label{eq:perturbation}
\end{equation}
To the zero-th order of $k$ we have ${\cal L}\left|P_{0}\right\rangle =0$
so $\left|P_{0}\right\rangle $ is the equilibrium distribution $\left|P_{\text{eq}}\right\rangle $
of the original process. To the first order in $k$ we have 
\[
\eta_{1}\left|P_{0}\right\rangle ={\cal L}\left|P_{1}\right\rangle +{\cal D}\left|P_{0}\right\rangle 
\]
Applying $\left\langle 1\right|$ to both side of the above equation,
we obtain 
\[
\eta_{1}=\left\langle 1\right|{\cal D}\left|P_{0}\right\rangle =\left\langle n_{\text{eq}}\right\rangle 
\]
which, in other words, is the well known relation $\left.d\lambda(k)/dk\right|_{k=0}=\left\langle n_{\text{eq}}\right\rangle $. 

More generally, defining the ``PGF$k$'' function $\phi$ as (and
omitting to write the variable $k$ explicitly)
\[
\phi(z,t)=\left\langle z^{n}|P\right\rangle =\left\langle z^{n}\right\rangle 
\]
its evolution is given by
\begin{eqnarray}
\frac{\partial\phi(z,t)}{\partial t} & = & \left\langle z^{n}|{\cal L}|P\right\rangle +k\left\langle z^{n}|{\cal D}|P\right\rangle \nonumber \\
 & = & (z-1)\left\langle z^{n}W^{+}(n)\right\rangle +(1/z-1)\left\langle z^{n}W^{-}(n)\right\rangle \nonumber \\
 & + & kz\frac{\partial\phi}{\partial z}\label{eq:dPGFk}
\end{eqnarray}
We see that compared to the original PGF, the evolution of the PGF$k$
adds only \emph{one} first derivative to the original evolution equation.
If we are able to compute the evolution of the PGF$k$ function, we
automatically possess the largest eigenvalue of ${\cal L}(k)$. As
we are only interested in the \emph{rate }of long time evolution,
in some cases as illustrated below, we don't even need to solve the
equation and we can restrict the investigation to some particular
points, as we will discuss in the next section.

Note that we could \emph{approximate} a jump process by a continuous
Fokker Plank (FP) equation
\[
\frac{\partial p}{\partial t}=-N\frac{\partial\left(ap\right)}{\partial x}+\frac{1}{2}\frac{\partial^{2}\left(bx\right)}{\partial x^{2}}
\]
where $N$ is a natural scale of the problem used for discretization,
$x=n/N$, $a(x)=\left(W^{+}(n)-W^{-}(n)\right)/N$, $b(x)=\left(W^{+}(n)+W^{-}(n)\right)/N$,
and follow the classical tilted backward operator method reviewed
by Touchette\citep{touchette2018introduction}. There are two disadvantages
with this method : first, for discrete jump processes, the Fokker
Plank approximation is of $O(1/N)$, which is a bad approximation
if $N$ is not large ; second, the term $b(x)$ is usually not a constant
and makes the backward, tilted FP operator rather intricate to investigate.

In the following, we illustrate the use of dPGF$k$ method through
few simple cases and investigate some of its extensions.

\section{Applications and extensions\label{sec:Applications-and-extensions}}

\subsection{Simple chemical reactions.\label{subsec:Simple-chemical-rates.}}

Consider a simple chemical reaction $\emptyset\leftrightarrows A$
which models for example the production of RNA when a gene is active\citep{paulsson2005modelsof}.
Denoting by $n$ the number of $A$ molecules, the transition rates
are 
\begin{equation}
W^{+}(n)=N\,\,\,;\,\,\,W^{-}(n)=n\label{eq:RNA:rates}
\end{equation}
Where the parameter $N$ (not necessarily an integer) is the production
rate of $A$. The evolution of the PGF$k$ is given by 
\begin{equation}
\frac{\partial\phi}{\partial t}=\left((k-1)z+1\right)\frac{\partial\phi}{\partial z}+N(z-1)\phi\label{eq:chemical:dPGFk}
\end{equation}
We can, if needed, eliminate $N$ from the equation by setting $\phi=\exp(Nu)$.

At the point $z^{*}=1/(1-k)$, the prefactor of the $\partial_{z}\phi$
in equation (\ref{eq:chemical:dPGFk}) is zero ; therefore, at this
point, $\phi(z^{*},t)$ evolves exponentially with rate 
\begin{equation}
\lambda(k)=N(z^{*}-1)=N\frac{k}{1-k}\label{eq:chemical:lam}
\end{equation}
The Legendre transform of the above relation is 
\begin{equation}
I(x)=\left(\sqrt{N}-\sqrt{x}\right)^{2}\label{eq:chemical:rate}
\end{equation}
This result has been obtained recently \citep{zilber2019agiant} by
using a WKB method.

If needed, we can check the above result by solving exactly equation
(\ref{eq:chemical:dPGFk}) using the methods of characteristics :
setting $\phi=\exp(Nu)$, the solution, for the initial condition
$\phi(z,t=0)=1$ is :
\[
u(z,t)=\frac{e^{(k-1)t}-1}{(k-1)^{2}}\left((k-1)z+1\right)+\frac{k}{1-k}t
\]
The linear term in $t$ (up to the scaling factor $N)$ is indeed
given by expression (\ref{eq:chemical:lam}). 

\subsection{The Ehrenfest urn.\label{subsec:The-Ehrenfest-urn.}}

The Ehrenfest urn is one of the first simple stochastic processes
used to understand the march toward equilibrium in statistical physics.
The large deviation of its long-time average was investigated recently
by Meerson and Zilber using a direct DV approach\citep{meerson2018largedeviations}.
In this model, $N$ objects are distributed among two urns ; at exponentially
distributed times, an object is drawn at random to change urn. Let
$n$ designates the size of the first urn, then the transition rates
for $n$ (up to a constant ) are given by: 

\begin{equation}
W^{+}(n)=N-n\,\,\,;\,\,\,W^{-}(n)=n\label{eq:Ehrenfest:rates}
\end{equation}
According to relation (\ref{eq:dPGFk}), the evolution of the PGF$k$
is given by 
\begin{equation}
\frac{\partial\phi}{\partial t}=N(z-1)\phi-(z^{2}-kz-1)\frac{\partial\phi}{\partial z}\label{eq:Ehrenefest:dPGF}
\end{equation}
which is a first order PDE. The system size $N$ can be eliminated
by setting $\phi=u^{N}$, which transforms equation (\ref{eq:Ehrenefest:dPGF})
into 
\begin{equation}
\frac{\partial u}{\partial t}+(z^{2}-kz-1)\frac{\partial u}{\partial z}=(z-1)u\label{eq:transformed1}
\end{equation}
Consider $z_{\pm}$, the two roots of the algebraic equation
\[
z^{2}-kz-1=0
\]
where $z_{+}$ is the positive one. At $z=z_{\pm}$, the prefactor
of $\partial_{z}u$ in equation (\ref{eq:transformed1}) vanishes
and we have 
\[
\frac{\partial u(z_{\pm},t)}{\partial t}=\left(z_{\pm}-1\right)u(z_{\pm},t)
\]
As $z_{+}>z_{-}$, the largest eigenvalue of ${\cal L}(k)$ is simply
\begin{eqnarray}
\lambda(k) & = & N(z_{+}-1)\nonumber \\
 & = & \frac{N}{2}\left(k-2+\sqrt{k^{2}+4}\right)\label{eq:Ehrenfest:lam}
\end{eqnarray}
The fixed point method allows us to avoid solving the partial differential
equation (\ref{eq:transformed1}) ; however, as this is a first order
linear PDE, it can be exactly solved. The solution of equation (\ref{eq:transformed1})
for the initial condition $u(z,0)=1$ is 
\[
u(z,t)=\left((z-z_{-})e^{\lambda_{+}t}-(z-z_{+})e^{\lambda_{-}t}\right)/(z_{+}-z_{-})
\]
where $\lambda_{\pm}=(z_{\pm}-1)$. Obviously, expression (\ref{eq:Ehrenfest:lam})
is indeed the correct largest eigenvalue.

The Legendre transform of expression (\ref{eq:Ehrenfest:lam}) is
\begin{equation}
I(x)=\left(N-2\sqrt{x(N-x)}\right)=\left(\sqrt{x}-\sqrt{N-x}\right)^{2}\label{eq:Ehrenfest:rate}
\end{equation}
Expression (\ref{eq:Ehrenfest:lam},\ref{eq:Ehrenfest:rate}) were
obtained by Meerson and Zilber \citep{meerson2018largedeviations}
using a direct DV approach.

The results of the above two subsections can be generalized. It can
be shown by elementary algebra (see appendix \ref{sec:First-order-polynomial}),
that when rates are first order polynomials in $n$, 
\begin{equation}
I(x)=\left(\sqrt{W^{+}(x)}-\sqrt{W^{-}(x)}\right)^{2}\label{eq:generalI}
\end{equation}
this expression has been obtained by other methods in \citep{zilber2019agiant}.

We stress that this expression is only correct for \emph{single} step
processes with \emph{first} \emph{order} polynomial rates. In general,
$I(x)$ must vanish for $x=\left\langle n_{\text{eq}}\right\rangle $
; In expression (\ref{eq:generalI}) however, $I(x)$ vanishes at
$x^{*}$ such that $W^{+}(x^{*})=W^{-}(x^{*})$. In general, $x^{*}\ne\left\langle n_{\text{eq}}\right\rangle $
and these two quantities coincide only for first order polynomial
rates of single step processes. 

\subsection{Extension to multi-step processes.}

\begin{figure}
\begin{centering}
\includegraphics[width=0.8\columnwidth]{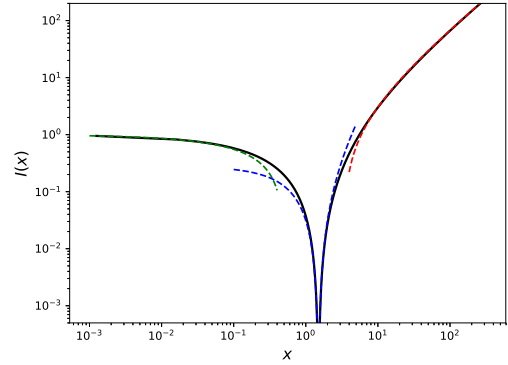}
\par\end{centering}
\caption{The rate function $I(x)$ for a two step process (eqs. \ref{eq:twosteps:rate1},\ref{eq:twosteps:rate2})
with $\alpha_{1}=\alpha_{2}=0.5$. Solid line : numerical Legendre
transform of $\lambda(k)$ (eq. \ref{eq:burst:lam}) ; dashed lines
: approximate expressions given by eqs (\ref{eq:burst:I0}-\ref{eq:burst:I2}).\label{fig:I:twosteps}}

\end{figure}

The dPGF$k$ method can easily be extended to investigate multi-step
processes. As an illustration, consider a simple generalization of
the chemical process considered in subsection \ref{subsec:Simple-chemical-rates.},
describing now RNA productions with \emph{bursts} \citep{golding2005realtime}
or the dynamics of neutron production in nuclear reactors\citep{houchmandzadeh2015neutron}.
The transition rates are :
\begin{eqnarray}
W(n\rightarrow n+m) & = & W_{m}^{+}(n)=N\alpha_{m}\label{eq:twosteps:rate1}\\
W(n\rightarrow n-1) & = & n\label{eq:twosteps:rate2}
\end{eqnarray}
where $\sum_{m=1}^{M}\alpha_{m}=1$. The coefficient $\alpha_{m}$
is the probability that a production events produces $m$ particles.
The previous case (equation \ref{eq:RNA:rates}) is recovered by setting
$\alpha_{1}=1$; as before, the parameter $N$ denotes the production
rate. Setting $\phi(z,t)=\exp\left(Nu(z,t)\right)$, the evolution
of the exponential part of PGF$k$ is given by (see appendix \ref{sec:The-dPGFk-equation.})
\begin{equation}
\frac{\partial u}{\partial t}=\left((k-1)z+1\right)\frac{\partial u}{\partial z}+\sum_{m=1}^{M}\alpha_{m}(z^{m}-1)\label{eq:burst:dPGFk}
\end{equation}
At the point $z^{*}=1/(1-k)$, the first order derivative in $z$
vanishes and therefore, 
\begin{equation}
\lambda(k)=\sum_{m=1}^{M}\frac{\alpha_{m}}{(1-k)^{m}}-1\label{eq:burst:lam}
\end{equation}
The above expression is not in general amenable to an analytic Legendre
transform, but is easily computed numerically. Moreover, considering
the limits $k\rightarrow-\infty$, $k\approx0$ and $k\rightarrow1$,
we can obtain the limiting form of $I(x)$ for $x\rightarrow0$ ,
$x\approx x^{*}$ and $x\rightarrow\infty$ :
\begin{eqnarray}
I(x) & \approx & 1-2\sqrt{\alpha_{1}x}\,\,\,\,\,\,\,\,\,\,\,x\ll x^{*}\label{eq:burst:I0}\\
I(x) & \approx & \frac{1}{2}\frac{(x-x^{*})^{2}}{\tilde{\alpha}}\,\,\,\,\,\,\,\,\,\,\,x\approx x^{*}\label{eq:burst:I1}\\
I(x) & \approx & x-(M+1)\alpha_{M}\left(\frac{x}{M\alpha_{M}}\right)^{\frac{M}{M+1}}\,\,\,x\gg x^{*}\label{eq:burst:I2}
\end{eqnarray}
where $x^{*}=\sum_{m}m\alpha_{m}$, $\tilde{\alpha}=\sum_{m}m(m+1)\alpha_{m}$
and we have supposed $\alpha_{1}\ne0$. Figure \ref{fig:I:twosteps}
illustrates the above results.

\subsection{Extension to multi-component systems.}

The dPGF$k$ method naturally generalizes to multi-component systems.
As an example, consider the simplest \citep{berg1978amodel,paulsson2005modelsof}
chemical reaction modeling the production of mRNA and its encoded
protein : 
\begin{eqnarray*}
W(n,p\rightarrow n+1,p) & = & N\\
W(n,p\rightarrow n-1,p) & = & n\\
W(n,p\rightarrow n,p+1) & = & \alpha n\\
W(n,p\rightarrow n,p-1) & = & \beta p
\end{eqnarray*}
where $(n,p)$ is the number of mRNA and proteins, $N$ the production
rate of mRNA, $\alpha,\beta$ the production and degradation rate
of proteins. Following the same arguments as above, the dPGFk equation
is 
\begin{eqnarray}
\frac{\partial\phi}{\partial t} & = & N(z_{1}-1)\phi+a(z_{1},z_{2})\frac{\partial\phi}{\partial z_{1}}+b(z_{1},z_{2})\frac{\partial\phi}{\partial z_{2}}\label{eq:multicomp}
\end{eqnarray}
where $z_{1},z_{2}$ are the conjugate variables to $n,p$, 
\begin{eqnarray}
a(z_{1},z_{2}) & = & (1-z_{1})+\alpha z_{1}(z_{2}-1)+k_{1}z_{1}\label{eq:fixa}\\
b(z_{1},z_{2}) & = & \beta(1-z_{2})+k_{2}z_{2}\label{eq:fixb}
\end{eqnarray}
and $k_{1},k_{2}$ are the amplitude of the tilted operator. Relation
(\ref{eq:multicomp}) is a linear first order PDE and can be solved
exactly ; as we are only interested in the rates, we can look as before
for vanishing points of the derivatives 
\[
a(z_{1}^{*},z_{2}^{*})=b(z_{1}^{*},z_{2}^{*})=0
\]
that is 
\begin{eqnarray*}
z_{2}^{*} & = & \frac{\beta}{\beta-k_{2}}\\
z_{1}^{*} & = & \frac{1}{1-k_{1}-\alpha(z_{2}^{*}-1)}
\end{eqnarray*}
and therefore 
\[
\lambda(k_{1},k_{2})=N(z_{1}^{*}-1)=\frac{k_{1}+\alpha(z_{2}^{*}-1)}{1-k_{1}-\alpha(z_{2}^{*}-1)}
\]
Taking the Legendre transform $I(\mathbf{x})=\sup_{\mathbf{k}}\left\{ \mathbf{k.x}-\lambda(\mathbf{k})\right\} $,
we find
\[
I(x_{1},x_{2})=\left(\sqrt{N}-\sqrt{x_{1}}\right)^{2}+\left(\sqrt{\alpha x_{1}}-\sqrt{\beta x_{2}}\right)^{2}
\]
We could have expected this result, as the transition rates for protein
production has the same form that those for mRNA production, where
$N$ has been replaced by $\alpha n$ : mRNA drives protein production
but in this simple scheme, its own production is independent of the
protein level. 

\subsection{Numerical computation.\label{subsec:Numerical-computation.}}

The dPGF$k$ method is well suited to compute numerically $\lambda(k)$.
As time flows, the solution of the dPGFk $\phi(z,t)$ converges to
$\text{\ensuremath{\phi_{0}(z)\exp\left(\lambda(k)t\right)}}$. The
dPGF$k$ equation is first order in time. Therefore, we can implement
a discrete numerical scheme on a finite suitable interval $[L_{0},L_{1}]$:
noting $dz,dt$ the discretization steps in in $z$ and $t$, $z_{i}=L_{0}+idz$
and $t_{j}=jdt$, one can compute $\phi(z_{i},t_{j+1})$ from $\phi(z_{i},t_{j})$.
At a chosen point $z_{\alpha}$, at each time step $t_{j}$, the ratio
$r_{j}=\phi(z_{\alpha},t_{j+1})/\phi(z_{\alpha},t_{j})$ is computed
and $\phi(z_{i},t_{j+1})$ is normalized by this ratio: 
\begin{figure}
\begin{centering}
\includegraphics[width=0.8\columnwidth]{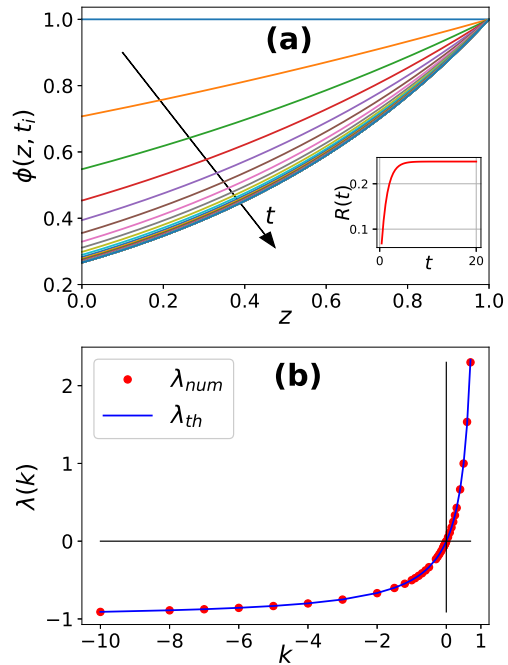}
\par\end{centering}
\caption{Numerical computation of $\lambda(k)$ of equation (\ref{eq:chemical:dPGFk})
with $N=1$ and initial condition $\phi(z,0)=1$. (a) For $k=0.25$,
evolution of $\phi(z,t)$ as a function of $z$ for 50 times $t_{i}\in[0,20]$.
At each time steps, $\phi(z,t+dt)$ is renormalized according to relation
(\ref{eq:phi_normalization}). Arrow indicates the direction of time.
Inset : $R(t)=\log\left(\phi(1,t+dt)/\phi(1,t)\right)/dt$ as a function
of time. (b)Numerical computation of $\lambda(k)$ as a function of
$k$ (circles) and compared to its theoretical value given by expression
(\ref{eq:chemical:lam}). \label{fig:Num-lam-k}}
\end{figure}
\begin{equation}
\phi(z_{i},t_{j+1})\leftarrow\phi(z_{i},t_{j+1})/r_{j}\label{eq:phi_normalization}
\end{equation}
With this normalization, $\phi(z_{i},t_{j})\rightarrow\phi_{0}(z_{i})$,
while 
\[
R_{j}=\frac{\log\left(r_{j}\right)}{dt}\rightarrow\lambda(k)
\]
This algorithm, which is similar in its principle to the Lanczos algorithm\citep{golub1996matrixcomputations},
is illustrated in figure \ref{fig:Num-lam-k} for the case of the
simple chemical reaction (equation \ref{eq:chemical:dPGFk}) discussed
in subsection \ref{subsec:Simple-chemical-rates.}.

\subsection{Discussions and conclusion.}

The application of the dPGF$k$ method we have presented in this article
was restricted to the cases where the transition rates $W(n\rightarrow n+m)$
were linear in the state of the system $n$. These systems give rise
to first order PDE for $\phi(z,t;k)$ and therefore are exactly soluble.
Moreover, as we are only interested in the long time behavior of the
function $\phi()$, we usually even do not  need to solve the PDE
but can restrict the analysis to some particular points $z^{*}$ where
the long time limit can be obtained through an ordinary differential
equation.

Many interesting stochastic processes are \emph{quadratic} in $n$
(see for example \citep{nemoto2014finitesize,zilber2019agiant} where
dynamical \emph{phase transition} are observed ). The dPGF$k$ method
for these cases leads to parabolic PDEs for $\phi()$ . The investigation
of these equations is beyond the scope of this article as there is
no general solution for them and they need to be investigated one
a case by case basis. These equations however are of Schrodinger type
and many more or less sophisticated methods are devoted in the literature
to their investigations.

Higher order rate transitions give rise to PDEs that are less studied
in the literature and therefore the dPGF$k$ method does not seem
to be very useful for their investigation, even though the numerical
method we have presented in subsection \ref{subsec:Numerical-computation.}
can still be used to obtain useful numerical results about their behavior.
On the other hand, for transition rates that are \emph{not }polynomial
in the state $n$, the dPGF$k$ method of this article seems to be
of limited use. 

More complicated quantities than the time-average, such as $\bar{f}=(1/T)\int_{0}^{T}f\left(n(t)\right)dt$
can also be considered with the dPGF$k$ method, with the same limitations
as discussed above. The diagonal matrix of DV in this case is ${\cal D}_{m}^{n}=\delta_{m}^{n}f(n)$
\citep{touchette2018introduction}; if $f(n)=n^{2}$, the dPGF$k$
method will contain a seconder order derivative of the form $z^{2}\partial^{2}\phi/\partial z^{2}$
and the resulting equation is still parabolic. Higher order terms,
as discussed above, would be more difficult to investigate. 

To summarize, in this paper, we have proposed a method (dPGF$k$)
to compute the rate function $I(x)$ by embedding the tilted matrix
of Donsker-Varadhan procedure into a partial differential equation,
obtaining its largest eigenvalue $\lambda(k)$ through the long time
analysis of the resulting equation and finding $I(x)$ by a Legendre
transform of $\lambda(k)$. We believe that this method can constitute
a useful tool in the analysis of large deviation of time-averaged
stochastic processes. 

\appendix

\section{The dPGF$k$ equation.\label{sec:The-dPGFk-equation.}}

The algebra for deriving the dPGF$k$ equation from the rate transitions
is straightforward. Defining the PGF as 
\[
\phi(z,t)=\sum_{n}z^{n}P(n,t)
\]
we have 
\begin{eqnarray*}
\partial_{t}\phi & = & \sum_{n}z^{n}\partial_{t}P(n,t)\\
\partial_{z}\phi & = & \sum_{n}nz^{n-1}P(n,t)\\
\partial_{zz}\phi & = & \sum_{n}n(n-1)z^{n-2}P(n,t)
\end{eqnarray*}
and so forth. Therefore, in the general dPGF$k$ equation : 
\begin{eqnarray}
\frac{\partial\phi(z,t)}{\partial t} & = & (z-1)\left\langle z^{n}W^{+}(n)\right\rangle +(1/z-1)\left\langle z^{n}W^{-}(n)\right\rangle \nonumber \\
 & + & kz\frac{\partial\phi}{\partial z}\label{eq:dPGFk:onestep}
\end{eqnarray}
a term such as $\left\langle az^{n}\right\rangle $ where $a$ is
a constant transforms into $\phi$, while a term such as $\left\langle nz^{n}\right\rangle $
transforms into $z\partial_{z}\phi$ etc.

For example, for $W^{+}(n)=N$ and $W^{-}(n)=n$ give rise to the
equation 
\[
\partial_{t}\phi=N(z-1)\phi+(1/z-1)z\partial_{z}\phi+kz\partial_{z}\phi
\]
which is equation (\ref{eq:chemical:dPGFk}) of subsection \ref{subsec:Simple-chemical-rates.}. 

For $W^{+}(n)=N-n$, $W^{-}(n)=n$ of subsection \ref{subsec:The-Ehrenfest-urn.},
we have 
\[
\partial_{t}\phi=(z-1)\left(N\phi-z\partial_{z}\phi\right)+(1/z-1)z\partial_{z}\phi+kz\partial_{z}\phi
\]
which is equation (\ref{subsec:The-Ehrenfest-urn.}). 

For multi-step processes with rates $W(n\rightarrow n+m)$, the dPGF$k$
equation is generally written as 
\[
\partial_{t}\phi=\sum_{m}(z^{m}-1)\left\langle z^{n}W(n\rightarrow n+m)\right\rangle +kz\partial_{z}\phi
\]
the rules for transforming $\left\langle ...\right\rangle $ into
derivatives of $\phi$ being the same.

\section{General expression of the rate function for linear jump rates\label{sec:First-order-polynomial}}

It has been mentioned in the discussions above that when the transition
rates $W(n\rightarrow n\pm1)$ are first order polynomials of $n$,
then the rate function is 
\[
I(x)=\left(W^{+}(x)-W^{-}(x)\right)^{2}
\]
We demonstrate this statement here.

Consider a one-step stochastic process 
\[
W^{+}(n)=a+bn\,\,\,;\,\,\,W^{-}(n)=c+dn
\]
the PGF$k$ associated to this process is 
\[
\frac{\partial\phi}{\partial t}=\left(bz^{2}-(b+d-k)z+d\right)\frac{\partial\phi}{\partial z}+\left(az+\frac{c}{z}-a-c\right)\phi
\]
Let us first consider the case $b=0$. The prefactor of $\partial_{z}\phi$
vanishes at 
\[
z^{*}=\frac{d}{d-k}
\]
and therefore 
\begin{equation}
\lambda(k)=az^{*}+\frac{c}{z^{*}}-a-c\label{eq:lambda:k}
\end{equation}
As $dz^{*}/dk=z^{*2}/d$, we have 
\[
x=\frac{d\lambda}{dk}=(az^{*2}-c)/d
\]
reversing this relation, we have
\[
z^{*2}=\frac{dx+c}{a}=\frac{W^{-}(x)}{W^{+}(x)}
\]
As $k=d(1-1/z^{*})$, the rates read: 
\begin{eqnarray}
I(x) & = & kx-\lambda=a(z^{*}-1)^{2}\nonumber \\
 & = & \left(\sqrt{W^{+}(x)}-\sqrt{W^{-}(x)}\right)^{2}\label{eq:ratefunction:general}
\end{eqnarray}
Consider now the case $b\ne0$ ; without loss of generality, we set
$b=1$. The prefactor vanishes for the roots $z_{\pm}$ of the second
order equation
\[
z^{2}-(1+d-k)z+d=0
\]
$\lambda(k)$ is still given by equation (\ref{eq:lambda:k}) and
\[
x=\frac{d\lambda}{dk}=\frac{-az_{+}+cz_{-}/d}{z_{+}-z_{-}}
\]
it is straightforward to show that, as before, $z_{+}^{2}=W^{-}(x)/W^{+}(x)$
and $z_{-}=d/z_{+}$. As $k=1+d-z_{+}-z_{-}$, after some algebraic
manipulation, we find
\begin{eqnarray*}
I(x) & = & kx-\lambda\\
 & = & W^{+}(x)+W^{-}(x)\pm2\sqrt{W^{+}(x)W^{-}(x)}
\end{eqnarray*}
As the rates $W$ are positive, we must choose the minus sign to have
$I(x^{*})=0$ where $x^{*}$is such that $W^{+}(x^{*})=W^{-}(x^{*})$.

\section{Numerical computations.}

The algorithm used for numerical computation of the largest eigenvalue
discussed in subsection \ref{subsec:Numerical-computation.} is an
explicit finite difference scheme written in C where the function
$\phi$ is discretized in space over $M=200$ points, $dz=\Delta L/M$
and $dt=0.5dz$. The algorithm for computing the discrete Legendre
transform follows directly the definition (\ref{eq:legendre-def})
and has been written in Julia language \citep{bezanson2017juliaa}. 

\paragraph*{Acknowledgments.}

We are grateful to Eric Bertin and Vivien Lecomte for detailed reading
of the manuscript and fruitful discussions.

\bibliographystyle{unsrt}

\end{document}